\documentclass[12pt]{article}
\newcommand{\ba}{\begin{eqnarray}}
\newcommand{\ea}{\end{eqnarray}}

\title{Dynamical Domain Wall Defects in $2+1$ Dimensions}
\author{C.D.\ Fosco$^1$\thanks{fosco@cab.cnea.edu.ar},
E.\ Fradkin$^2$ \thanks{eduardo@buenosaires.physics.uiuc.edu} and A.
L\'opez$^1$\thanks{lopez@cabtep2.cnea.edu.ar}\\
 {\normalsize \it $^1$Centro
At\'omico Bariloche, 8400 Bariloche, Argentina} \\  {\normalsize \it
$^2$Department of Physics, University of Illinois at Urbana-\-Champaign,}\\
{\normalsize \it 1110 West Green Street,Urbana, IL 61801-3080} }

\begin{document}
\maketitle

\begin{abstract}
We study some dynamical properties of a Dirac field in $2+1$ dimensions
with spacetime dependent domain wall defects.
We show that the Callan and Harvey mechanism applies even to the case
of defects of arbitrary shape, and in a general state of motion.
The resulting chiral
zero modes are localized on the worldsheet of the defect, an embedded
curved two dimensional manifold. The dynamics of these zero modes is
governed by the corresponding induced metric and spin connection. Using
known results about determinants and anomalies for fermions on surfaces
embedded in higher dimensional spacetimes, we show that the chiral anomaly
for this two dimensional theory is responsible for the generation of a
current along the defect. We derive the general expression for
such a current in terms of the geometry of the defect, and show that
it  may be interpreted as due to an
``inertial" electric field, which can be expressed entirely in terms
of the spacetime curvature of the defects.
We discuss the application of this
framework to fermionic systems with defects in condensed matter.
\end{abstract}
\bigskip
\newpage

The behavior of fermions in the background of defects is a problem of
general interest in many areas of physics ranging from the behavior of
textures in superfluid phases of $He_3$ \cite{he3} to cosmic 
strings~\cite{cosmic}. The prototype of these problems is the Callan-Harvey
effect \cite{callan} in which a Fermi field (whose physical meaning
depends on the specific problem) in an odd-dimensional spacetime couples 
to the defect through a mass term  which takes a positive value on half of
the space and a negative value on the other half. Under these
circumstances, the matter (Fermi) field has a bound state (a zero mode)
at the defect which is capable of carrying currents. Thus, the presence
of the defect leads to an anomaly in the gauge current in the bulk which
is exactly compensated by an equal and opposite anomaly at the defect.
A similar phenomenon is also found at the edge of a droplet of a
(non-relativistic) two-dimensional electron gas in a magnetic field in
the regime of the fractional and integer quantum Hall effects,
where the anomaly in the bulk Hall current is cancelled by the gauge anomaly of
the chiral edge states\cite{wen,haldane}.

While in most problems static defects are sufficient to describe the
physics, in many others (particularly in condensed matter) it is of
current interest to understand the possible effects due to the {\sl dynamics} 
of the defects. In particular, it is interesting to investigate
the interaction of the fermion zero modes with
the fluctuating geometry induced by the dynamics of the defect.

In this letter we shall generalize the results of \cite{fl}, which dealt
with a Dirac field in $2+1$ dimensions having space dependent domain wall
like defects in their mass, to the case of an arbitrary, space {\em and
time\/} dependent defect. We shall show that the analog of the Callan and
Harvey mechanism~\cite{callan}, namely, the existence of a chiral zero
mode, still holds, with the zero mode now localized on the defect
worldsheet. An important difference is that now, even in the absence of any
external gauge field, there is an induced fermionic current on the defect,
caused by the gravitational chiral anomaly for the fermions localized on
the defect. This current can be thought of as due to the existence of an
effective or ``inertial" electric field, defined in terms of the geometry of
the defect (the term ``inertial" comes from the fact that it vanishes when
the defect is not accelerated).

The action for a Dirac fermion in three Euclidean dimensions, described by
the Cartesian coordinates $x^0,x^1,x^2$ ($x^0$ denoting the Euclidean
time), in the presence of a space-time dependent mass $M(x)$, is given by
\begin{equation} \label{actn}
    S \;=\;\int d^3 x \, {\bar\Psi}(x)\,[\gamma^{\alpha}D_{\alpha} + M
(x)]\,\Psi (x)
\end{equation}
where $D_{\mu}= \partial _{\mu} + i e A_{\mu}$. We have included a gauge
field for the sake of generality, although it shall be taken to be zero in
most of this work. The Euclidean Dirac matrices $\gamma^\alpha$ are chosen
according to
\begin{equation}\label{dfgp}
\gamma^0\;=\;\sigma^3 \;\;\;\gamma^1\;=\;\sigma^1
\;\;\;\gamma^2\;=\;\sigma^2\;,
\end{equation}
where $\sigma^\alpha$ denote the usual Pauli matrices. We adopt the
convention of using letters from the first part of the Greek alphabet to
denote tensor components with respect to Cartesian coordinates. Moreover,
as the three-dimensional space-time is assumed to be flat, these indices
will be raised and lowered by using the identity tensor
$\delta_{\alpha\beta}=\delta^{\alpha\beta}$.  Letters from the end of the
Greek alphabet shall instead be reserved for tensor components in general
coordinates.

Rather than considering arbitrary configurations for the mass, we
shall concentrate on those having a spacetime dependent step-like
defect.
This kind of configuration is conveniently parametrized as
\begin{equation}\label{mass}
    M(x) = \Lambda \; sign [\varphi(x)]
\end{equation}
where $\varphi(x)= \varphi(x^{\alpha})$ is a scalar field changing sign
along a surface which, by definition, is the worldsheet of the defect.
$\Lambda$ is a constant with the dimensions of a mass, characterizing (half
of) the height of the defect. Since the defect is completely determined by
the behaviour of $\varphi$ in an infinitesimal neighborhood of that
surface, we are entitled to extend its definition in the most convenient
way, to cover all the spacetime.

Here we will consider the simple case of
defects defined by scalar fields $\varphi(x)$ with the
property that $\varphi$ are solutions of Laplace's equation, and hence
the surfaces of constant $\varphi$ split $2+1$-dimensional spacetime in
two. In other terms,
$\varphi$ defines a foliation of spacetime,
namely, if the three-dimensional spacetime considered is a manifold ${\cal
M}$ (typically, ${\cal M}=\Re^3$), then the following relations hold:
\ba
&& {\cal M} = {\bigcup}_{c \in \Re} \;  \{ x: \varphi(x)=c \}  \nonumber\\
&& \{ x: \varphi(x)=c \} \bigcap \{ x: \varphi(x)=c' \} = \emptyset \; ,
\forall c\neq c' \in \Re \,. \label{fo}
\ea
These condition are of course not sufficient to uniquely determine the
function $\varphi$. We shall see below that some other constraints,
compatible with (\ref{fo}), arise naturally when seeking for the simplest
analytical description of the geometry of the system.  We first need to
write the three-dimensional action in general coordinates $x'^{\mu}$
\begin{equation}\label{actgen}
    S \;=\;\int d^3 x' \; g^{\frac{1}{2}} \; {\bar\Psi} \,[\gamma^{\mu}
D_{\mu} + M (\varphi)] \,\Psi
\end{equation}
where
\begin{equation}\label{dfgc}
    \gamma^\mu \;=\; \frac{\partial x'^\mu}{\partial x^\alpha}\, \gamma^\alpha
\;\Rightarrow\, \{ \gamma^\mu , \gamma^\nu \} \;=\; 2 \,g^{\mu\nu}
\end{equation}
with $g_{\mu\nu}$ denoting the metric in the new coordinates, namely
\begin{equation}\label{dfgmn}
    g_{\mu\nu} \;=\; \delta_{\alpha\beta} \frac{\partial x^\alpha}{\partial
x'^\mu}  \frac{\partial x^\beta}{\partial x'^\nu} \;.
\end{equation}

We now impose some requirements on the (choice of) coordinates $x'$, in
order to have a simple description of the zero mode, and to disentangle (as
far as possible) its dynamics from the one of the bulk degrees of freedom.
These come from the fermionic modes defined in the region where the mass
has a finite value. We define one of the new coordinates as $\varphi$. This
choice singles out a surface for each value of $\varphi$, the defect
corresponding to $\varphi = 0$. The other two coordinates define constant
$\varphi$ surfaces in parametric form. They are then orthogonal to
$\varphi$, what implies the general expression for the length element:
\begin{equation}
    ds^2 = h_{\varphi}^2({\xi},\varphi) d\varphi^2 + g_{ab}({\xi},{\varphi})
    d\xi^a d\xi^b  \label{cc}
\end{equation}
where $\xi^0= x'^0$, and $\xi^1= x'^1$. We introduced the special notation
$\xi^a,\, a=0,1$ for two of the coordinates, since they shall parametrize
the constant-$\varphi$ surfaces, and will be the proper coordinates on a
curved two dimensional spacetime when studying the chiral zero mode
dynamics. Namely, the defect will be expressed in parametric form as
\begin{equation}
  X^\alpha \;=\;  X^\alpha (\xi^0,\xi^1) \;, \alpha\,=\,0,1,2.
\end{equation}

It is always possible, by a redefinition of $\varphi$, to have $h_{\varphi}
=1$. In fact, this amounts to redefining $\varphi$ in such a way that it
still defines a foliation, and moreover, if one moves an amount $d \varphi$
in the normal direction to a surface of constant $\varphi$ the length of
such displacement is just $d \varphi$. We assume in what follows that such
a choice of $\varphi$ has been made. Then
\begin{equation}
   ds^2 = d\varphi^2 + g_{ab}(\xi,{\varphi}) d\xi^a d\xi^b
\end{equation}
where
\begin{equation}
   g_{ab}= {\frac {\partial x^{\alpha}} {\partial
          \xi^{a}}}{\frac{\partial x^{\alpha}}{\partial \xi^{b}}} \;.
\end{equation}
In these coordinates, the Dirac matrices:
    $\gamma^a = \displaystyle{\frac{\partial \xi^a}{\partial x^\alpha}}
    \gamma^\alpha$, $\gamma_{\varphi} = \displaystyle{\frac{\partial
    \varphi}{\partial x^\alpha}}
    \gamma^\alpha$  satisfy the algebraic relations
\begin{equation}\label{acomm}
    \{\gamma ^a, \gamma^b \} \;=\; 2 \,g^{a b} \;\;,\;\;
    \{\gamma_{\varphi}, \gamma^a \} \;=\; 0 \;\;,\;\;
    (\gamma_{\varphi})^2 \;=\; 1 \;.
\end{equation}
We note that the two matrices $\gamma^a$ play the role of Dirac matrices
for the two dimensional manifold $\varphi = constant$, while
$\gamma_{\varphi}$ plays the role of a (chirality) $\gamma_5$ matrix, and
we shall adopt the notation $\gamma_5 \equiv \gamma_\varphi$. Indeed, the
latter is the projection of the Dirac matrices in three dimensions along
the direction of the unit normal vector to each surface,
\begin{equation}\label{gams}
  \gamma^5 (\xi,\varphi) \;=\; {\hat n}^\alpha (\xi,\varphi) \gamma^\alpha
\end{equation}
where
\begin{equation}\label{dfnor}
  n^\alpha (\xi,\varphi) \;=\;g^{-\frac{1}{2}}(\xi,\varphi)
  \epsilon^{\alpha\beta\lambda}\partial_0 X^\beta(\xi,\varphi) \,
  \partial_1 X^\lambda(\xi,\varphi)
\end{equation}
with $\epsilon^{\alpha\beta\lambda}$ the Euclidean Levi-Civita symbol, and
$g = \det(g_{ab})$.

We now write the action (\ref{actgen}) as
\begin{equation}\label{actgen2}
    S \;=\;\int d^2 \xi d\varphi \; g^{\frac{1}{2}} \; {\bar\Psi}(\xi,\varphi)
\,{\cal
    D}
     \,\Psi(\xi,\varphi)
\end{equation}
in terms of the operator ${\cal D}= \gamma^{\mu} D_{\mu} + M (\varphi)$
which in our coordinates reads
\begin{equation}
    {\cal D}= \gamma^5 D_\varphi + M (\varphi)+ \gamma^{a} D_{a}\;.
\end{equation}
Assuming now $A_\varphi=0$, $A_a= A_a (\xi)$ (or a configuration connected
to this one by a gauge transformation), we see that:
\begin{equation}\label{dec}
    {\cal D}= [\partial_\varphi + M (\varphi)+ \not \! d] {\cal P}_L
              +[-\partial_\varphi + M (\varphi)+ \not \! d] {\cal P}_R
              \;,
\end{equation}
where we introduced the projectors
\begin{equation}
    {\cal P}_L \;=\; \frac{1+\gamma^5}{2} \;\;\;,\;\;\;{\cal P}_R \;=\;
    \frac{1-\gamma^5}{2} \;.
\end{equation}
and $\not \! d$ the two dimensional Euclidean Dirac operator
corresponding to the two coordinates $\xi^a$, namely
\begin{equation}\label{dfsd}
   \not \! d \;=\; \gamma^a \partial_a \;.
\end{equation}

We note that, even with this special choice of coordinates, both the
function $g^{\frac{1}{2}}$ and the operator $\not \!\! d$ will, in general,
depend on the coordinate $\varphi$. However, we shall assume that both, the
curvature radius of the defect in the spatial plane, and the characteristic
length of the fluctuations in its shape, are  much larger than its
localization length in the spatial plane. Therefore we may regard $\not \!
d$ and $g^{\frac{1}{2}}$ as $\varphi$-independent. This is obvious for the
modes that are concentrated near to the defect but it is, indeed, also true
for the case of non-localized modes, since they are sensitive only to the
module of the mass, which is assumed to be constant outside the defect.

We can now borrow, with slight modifications, the results from
\cite{fl} to rewrite the action (\ref{actgen2}) in terms of an
infinite number of two-dimensional actions, one of which shall
describe the zero mode. To that end we introduce $a$ and
$a^\dagger$, operators acting on functions of $\varphi$ only,
\begin{equation}\label{dfaad}
  a\;=\;\partial_\varphi + M(\varphi) \;\;,\;\;
  a^\dagger\;=\;-\partial_\varphi + M(\varphi) \;,
\end{equation}
so that
\begin{equation}\label{newd}
   {\cal D} \;=\; (a+\not\!d) {\cal P}_L  + (a^\dagger+\not\!d) {\cal P}_R
\end{equation}

We can extract the dependence on $\varphi$ from the fields, by expanding
them in the modes of the Hermitian operator built in terms of ${\cal D}$
and its adjoint (${\cal D}$ itself is not Hermitian)
\begin{equation}\label{defh}
   {\cal H} \;=\; {\cal D}^\dagger {\cal D} \;=\; (h - \not\!d^2)
   {\cal P}_L + ({\tilde h} - \not \! d^2) {\cal P}_R \;,
\end{equation}
with $h=a^\dagger a$ and ${\tilde h}=a a^\dagger$. The expansions
of the fermionic fields will read \ba\label{epsi} \Psi
(\varphi,\xi) &=& \sum_n \left[ \phi_n (\varphi) \psi^{(n)}_L
(\xi) +   {\tilde \phi}_n (\varphi) \psi^{(n)}_R (\xi) \right]
\nonumber\\ {\bar \Psi}(\varphi,\xi) &=& \sum_n \left[ {\bar
\psi}^{(n)}_L (\xi) \phi_n^\dagger (\varphi) +{\bar \psi}^{(n)}_R
(\xi) {\tilde \phi}_n^\dagger (\varphi) \right]
 \ea
where the subscripts $L, R$ denote eigenvectors of ${\cal P}_L$
and ${\cal P}_R$, respectively. The eigenfunctions $\phi_n$ and
${\tilde \phi}_n$ satisfy
 $$ h \phi_n (\varphi) \;=\; \lambda_n^2 \phi_n (\varphi)\;\;,\;\; {\tilde
h} {\tilde \phi}_n (\varphi) \;=\; \lambda_n^2 {\tilde \phi}_n (\varphi) $$
\begin{equation}\label{eephi}
    \langle \phi_n | \phi_m \rangle \;=\; \delta_{n,m} \;\;,\;\; \langle
    {\tilde \phi}_n | {\tilde \phi}_m \rangle \;=\; \delta_{n,m} \;.
\end{equation}

The operators $h$ and ${\tilde h}$ are positive (the $\lambda_n$ are
assumed to be real and we fix their sign, by convention, to be positive)
and have the same spectrum, with the only exception of the zero mode
$\lambda_n = 0$, since, for any $\phi_n$ with $\lambda_n \neq 0$, there
also exists one eigenvector of ${\tilde h}$ with identical eigenvalue
\begin{equation}
h \phi_n (\varphi) \;=\; \lambda_n^2 \phi_n (\varphi) \;\Rightarrow\;
{\tilde h} \; [\frac{1}{\lambda_n} a\phi_n (\varphi)] \;=\; \lambda_n^2 \;
[\frac{1}{\lambda_n} a\phi_n (\varphi)] \label{i23}
\end{equation}
where the factor $\frac{1}{\lambda_n}$ is introduced to normalize the
eigenvectors of ${\tilde h}$. (Of course the reciprocal property for the
eigenstates of ${\tilde h}$ also holds.)

We see that there is a zero mode for the operator $a$, whose explicit
expression is
\begin{equation}\label{zmd}
    \Psi_{L}^{(0)} (x) \;=\; e^{-\int^{\varphi} d{\tilde \varphi}
    M({\tilde \varphi})}   \chi (\xi)
\label{ze}
\end{equation}
where $\chi$ satisfies
\begin{equation}
    \gamma^a D_a {\cal P}_L \chi (\xi) =0 \;,
\end{equation}
namely, it is a chiral zero mode.

 Let us now discuss the contribution of
this mode to the effective action. The chiral zero mode, being a
massless excitation, will dominate the effective low-energy
dynamics, and it will be the only mode contributing  to the
current induced on the defect due to the fluctuations in its
shape.

Inserting the expansion (\ref{epsi}) into the action
(\ref{actgen2}), and keeping only the term corresponding to the
zero mode (\ref{zmd}), we see that the resulting action is the one
for a chiral fermion on a two dimensional curved Euclidean
manifold (the worldsheet of the defect), which, when put in the
symmetric form, becomes
\begin{equation}\label{twodact}
    S_0 \;=\; \frac{1}{2} \int \, d^2 \xi \, g^{\frac{1}{2}} \,
    [ {\bar\chi}(\xi)\,\gamma^a  {\cal P}_L \,\partial_a \chi(\xi)\,-\,
    \partial_a {\bar\chi}(\xi)\,\gamma^a  {\cal P}_L \,\chi(\xi)] \;.
\end{equation}
We have assumed that the $\varphi$ dependent localization factor is
normalized to one~\footnote{All the objects are of course evaluated at
${\varphi} = 0$.} and we have dropped the dependence on the gauge field.

Following reference \cite{KKS}, we rewrite this action as
\begin{equation}
  S_0 \;=\; \int \, d^2\xi \, g^{\frac{1}{2}} \,
    {\bar\chi}(\xi)\,\gamma^a(\partial_a + \Gamma_a )\,{\cal P}_L
    \chi(\xi)
\end{equation}
where
\begin{equation}
  \Gamma_a \,=\, \frac{1}{2} \gamma^b \nabla_a \gamma_b \,
             = \,\frac{1}{2}\gamma^b n^{\alpha} \partial_a
             \partial_b  X_\alpha.
\end{equation}

As shown in \cite{KKS}, the fermionic current has an anomaly
\begin{equation}\label{canom}
  \nabla_a J^a_{L} (\xi) \;=\; \partial_a  J _L ^a + \Gamma^a_{ba} J^b_L \,
                    =\, - \frac{i}{48 \pi} \, \sqrt{g} \, R
\end{equation}
where $\Gamma^a_{bc} =\frac{1}{2} g^{ad} (g_{db,c} + g_{dc,b}-g_{bc,d})$ is
the metric connection, in particular $\Gamma^a_{ba} = g^{-\frac{1}{2}}
\partial_b g^{\frac{1}{2}} $.
 $R$ is the scalar curvature,  which reads
\begin{equation}
R \;=\; - 2 g^{-1} \; R_{0101}
\end{equation}
where $R_{0 1 0 1}$ is the (only) independent component of the Riemmann
tensor. This can be written in terms of the induced metric and its
derivatives as follows

$$ R_{0101} \;=\; \frac{1}{2} \, \left( 2 g_{01,01} - g_{11,00} - g_{00,11}
\right) $$ $$ -\frac{1}{4} g^{a b} \left[ ( 2 g_{0b,0} - g_{00,b} ) ( 2
g_{1a,1} - g_{11,a} ) \,\right.$$
\begin{equation}
\left.-\, (g_{0b,1} + g_{1b,0} - g_{01,b}) (g_{1a,0} + g_{0a,1} -
g_{01,a})\right] \;.
\end{equation}
It is also convenient to have an expression for the curvature in terms of
derivatives of the parametric equations which define the defect space-time
surface:

$$ R \;=\; - 2 g^{-2} \, \left\{ \left\bracevert
\begin{array}{ccc}
  {\vec X}_{00} \cdot {\vec X}_{11} & {\vec X}_{00} \cdot {\vec X}_{0}  &
  {\vec X}_{00} \cdot {\vec X}_{1} \\
  {\vec X}_{0} \cdot {\vec X}_{11} & {\vec X}_0 \cdot {\vec X}_0 &
  {\vec X}_0 \cdot {\vec X} _1 \\
  {\vec X}_1 \cdot {\vec X} _{11} & {\vec X}_1 \cdot {\vec X}_0 &
  {\vec X}_1 \cdot {\vec X} _1
\end{array}
\right\bracevert \right. $$
\begin{equation}
\left. - \left\bracevert
\begin{array}{ccc}
  {\vec X}_{01} \cdot {\vec X}_{01} & {\vec X}_{01} \cdot {\vec X}_{0}  &
  {\vec X}_{01} \cdot {\vec X}_{1} \\
  {\vec X}_{0} \cdot {\vec X}_{01} & {\vec X}_0 \cdot {\vec X}_0 &
  {\vec X}_0 \cdot {\vec X} _1 \\
  {\vec X}_1 \cdot {\vec X} _{01} & {\vec X}_1 \cdot {\vec X}_0 &
  {\vec X}_1 \cdot {\vec X} _1
\end{array}
\right\bracevert \right\}
\end{equation}

 For the case of Dirac fermions on a flat $1+1$
dimensional spacetime coupled to an Abelian gauge field, the knowledge of
the chiral anomaly together with the conservation equation for the current,
enables one to determine the relation between the fermionic current and the
external field. We show here that in the case we are dealing with, the
gravitational chiral anomaly and the conservation equation for the current
also determine the form of the fermionic current, but now as a function
only of the curvature of the defect.

We first note that equation (\ref{canom}), together with the fact that the
vector current $J^a (\xi)$ is non-anomalous, implies the two relations:
\begin{eqnarray}\label{nonan}
    \nabla_a J^a (\xi) &=& g^ {-\frac{1}{2}}
    \partial_a ( g^ {\frac{1}{2}} J^a)= 0
    \nonumber \\
   \epsilon^{ab} \partial_a J_b &=& \partial_0 J_1 (\xi) -
   \partial_1 J_0 (\xi) \;=\;-\frac{i}{24\pi} \,
   \sqrt{g} \, R \;.
\end{eqnarray}

In two dimensions, the current can be written as the gradient of a scalar
field $\phi$ plus the curl of a pseudoscalar field $\sigma$,
\begin{equation}\label{cci}
    J_b (\xi) \;=\; \partial_b \phi + i g^{\frac{1}{2}}
     \epsilon_{bc}  \partial^c \sigma
\end{equation}
where the factor $g^{\frac{1}{2}}$ has been introduced in order to
compensate for the Jacobian factor due to the Levi-Civita symbol.
 Imposing conditions (\ref{nonan}) to the current expressed in terms of the
scalar fields (\ref{cci}), we deduce that the functions $\phi$ and $\sigma$
must satisfy
\begin{eqnarray}\label{csigma}
g^{-\frac{1}{2}}\partial_a ( g^{\frac{1}{2}} \partial^a \phi) &=& \Delta
\phi \;=\; 0 \nonumber \\
     g^{-\frac{1}{2}} \partial_a ( g^{\frac{1}{2}} \partial^a \sigma)
       &=& \Delta \sigma \;=\; -\frac{1}{24\pi}  \, R
\end{eqnarray}
where $\Delta$ denotes the Laplacian acting on scalar fields.
 Therefore, by solving (\ref{csigma}) for $\phi$ and $\sigma$ and
substituting it into (\ref{cc}) one can find the fermionic current in terms
of the curvature of the defect. Of course, the first equation implies that
the field $\phi$ vanishes, thus we arrive to the general expression

\begin{equation}\label{gener}
  J_a (\xi) \;=\; - \frac{i}{24\pi} \,
  g^{\frac{1}{2}} \,\epsilon_{ab}\partial^b
   \Delta^{-1} R \;.
\end{equation}

For the case of a defect whose configuration is close to a static
rectilinear wall, we shall obtain an approximate expression for
(\ref{gener}). We assume the configuration has the form
 \begin{equation}\label{flc1}
  X^\alpha (\xi^a) \;=\; {\bar X}^\alpha (\xi^a) \,+\,
  Y^\alpha (\xi^a)
 \end{equation}
where ${\bar X}^\alpha (\xi^a)$ denotes the static rectilinear defect which
is localized on, say, the $X^2 = 0$ plane
 \begin{equation}\label{flc2}
  {\bar X}^a (\xi^a) \;=\; \xi^a \; (a=0,1) \;\;\;,\;\;\;
  {\bar X}^2 (\xi^a)\;=\;0\;,
 \end{equation}
and $Y^\alpha (\xi^a)$ is the fluctuating part
 \begin{equation}\label{flc3}
Y^\alpha (\xi^a) \;=\; \delta^\alpha_2 \, \eta (\xi^a)
 \end{equation}
with $\eta$ a small but otherwise arbitrary function.

We then expand (\ref{gener}) in powers of $\eta$, which keeping only the
leading (quadratic) term yields, after some algebra
 \begin{equation}\label{pert}
   J_a (\xi) \;=\;  \frac{i}{12\pi} \,\epsilon_{ab}\partial^b
   \Delta_0^{-1} \det (\frac{\partial^2 \eta}{\partial_c\partial_d})\;,
 \end{equation}
where $\Delta_0 = \partial_a \partial_a$ is the {\em free\/} Laplacian. It
is interesting to notice the striking similarity between (\ref{pert}) and
the corresponding formula yielding the electric current $J^{el}$ for a
fermion on a flat two dimensional spacetime, but in the presence of an
external electric field $E$
 \begin{equation}\label{elec}
   J^{el}_a (\xi) \;=\;  \frac{i}{2\pi} \,\epsilon_{ab}\partial^b
   \Delta_0^{-1} E\;,
 \end{equation}
which allows we to define an effective geometric electric field $E_{eff}$
 \begin{equation}\label{efec}
  E_{eff}\;=\;  \frac{1}{6} \,
  \det (\frac{\partial^2 \eta}{\partial_c\partial_d})
 \;=\; \frac{1}{6} \left[\partial^2_0 \eta \,\partial^2_1 \eta -
        (\partial_0\partial_1\eta)^2  \right]\;.
 \end{equation}
In particular, for all the configurations such that $\eta$ is a quadratic
form in the coordinates, $\det (\frac{\partial^2
\eta}{\partial_c\partial_d})$ becomes a constant. That constant will
be non-zero whenever the quadratic form is non-degenerated, namely,
when the surface does indeed have a non-zero curvature. In this case
the induced current is the analogous of the one obtained by applying a
uniform electric field to a one dimensional system of massless Dirac
fermions.

Our result has, as we stated above, a simple and intuitive
explanation: as the defect moves, relative to the frame of the stationary
defect,
the fermions feel an ``inertial"  electric field which expresses the fact
that the fermions are being ``left behind" by the moving defect. On the
other hand the current induced by the motion of the defect, being anomalous,
has the interpretation of a local change of the zero of the energy of the Dirac
sea
({\it i.\ e.\/} the ``chemical potential").

It is worth to note that effects of
this type are not a peculiarity of Dirac fermions, and that they also
appear for
non-relativistic particles on a moving defect. In fact, an analog of this
problem
has been  considered recently in the context of formation of ``stripes" in
high-temperature superconductors and effects of this type were found
to play a significant role\cite{nature}. Finally, given that the mass
term in a Dirac field in $2+1$ dimensions breaks time reversal
invariance, it is natural to expect, not only  that similar effects should be
present in anisotropic states of the two-dimensional electron gas in
magnetic fields\cite{eisenstein}, but that a description involving a Dirac
field with
defects should play an important role in their theoretical description.

{\bf Acknowledgements}: This work begun during a visit of EF to
Instituto de F{\'\i}sica Balseiro and he thanks
A.\ Garc{\'\i}a from for his kind hospitality. We acknowledge F.A. Schaposnik 
for useful comments and references.
C.D.F. and A.L. thank the members of the Department of Physics of the
University of Oxford where part of this work was done.
This work was supported in part by CONICET, The British Council, 
Fundaci\'on Antorchas, and ANPCyT (Argentina) (CF and AL) and 
by the National Science Foundation
grant No.\ DMR98-17941 at the University of Illinois at Urbana-Champaign
(EF).

\newpage


\end{document}